\documentclass[12pt,preprint]{aastex}

\slugcomment{N}
\shorttitle{Earth rotation rate}
\shortauthors{Pall\'e et al., 2005}

\begin{document}

\title{Identifying the rotation rate and the presence of dynamic weather on extrasolar
Earth-like planets from photometric observations.}

\author{E. Pall\'e\altaffilmark{1}, Eric B. Ford\altaffilmark{2}, S. Seager\altaffilmark{3},
P. Monta\~n\'es-Rodr\'{\i}guez\altaffilmark{1}, M.
Vazquez\altaffilmark{1}}

\affil{Instituto de Astrofisica de Canarias, La Laguna, E38200,
Spain. }\email{epalle@iac.es, pmr@iac.es, mva@iac.es}

\affil{Department of Astronomy, University of Florida, 211 Bryant
Space Science Center, PO Box 112055 Gainesville, FL, 32611-2055,
USA}\email{eford@astro.ufl.edu}

\affil{EAPS, Massachusetts Institute of Technology, Cambridge,
MA02139-4307, USA }\email{seager@MIT.EDU}

\begin{abstract}

With the recent discoveries of hundreds of extrasolar planets, the
search for planets like Earth and life in the universe, is quickly
gaining momentum. In the future, large space observatories could
directly detect the light scattered from rocky planets, but they
would not be able to spatially resolve a planet's surface. Using
reflectance models and real cloud data from satellite observations,
here we show that, despite Earth's dynamic weather patterns, the
light scattered by the Earth to a hypothetical distant observer as a
function of time contains sufficient information to accurately
measure Earth's rotation period. This is because ocean currents and
continents result in relatively stable averaged global cloud
patterns. The accuracy of these measurements will vary with the
viewing geometry and other observational constraints. If the
rotation period can be measured with accuracy, data spanning several
months could be coherently combined to obtain spectroscopic
information about individual regions of the planetary surface.
Moreover, deviations from a periodic signal can be used to infer the
presence of relatively short-live structures in its atmosphere
(i.e., clouds). This could provide a useful technique for
recognizing exoplanets that have active weather systems, changing on
a timescale comparable to their rotation. Such variability is likely
to be related to the atmospheric temperature and pressure being near
a phase transition and could support the possibility of liquid water
on the planet's surface.

\end{abstract} 

\section{Introduction}

Over the past two decades, more than 240 planets have been
discovered orbiting stars other than the Sun. To date all planets
discovered around main sequence stars are significantly more massive
than the rocky planets of the solar system. Radial velocity surveys,
however, are starting to detect rocky planet candidates below 10
Earth masses (Rivera et al., 2006; Udry et al, 2007) and, for the
coming decades, ambitious space missions are being proposed that
would be able to detect nearby planets with physical properties
similar to Earth (see, e.g., Lindensmith , 2003; Kaltenegger, 2005;
Fridlund, 2004; Cash, 2005; Schneider et al, 2006).

Among other important physical properties, the identification of the
rotation rate of an exoplanet with relatively high accuracy will be
important for several reasons (Laskar and Correia, 2004). First,
measuring the rotation rate can help to understand the formation
mechanisms and dynamical evolution of extrasolar planetary systems
(Agnor et al., 1999; Chambers, 2001; Goldreich et al., 2004). For
example, are planetary rotation periods smoothly varying as a
function of the planet mass and semi-major axis, as would be
expected if the planet's angular momentum is dominated by the
gradual accretion of small planetesimals? Or are planet's rotation
periods essentially uncorrelated with their mass and orbital
properties, as would be the case if the planet's angular momentum is
dominated by the late accretion of a few large impactors?  The
rotation periods of a sample of planets could be directly compared
to numerical simulations of planetary formation that track the spin
evolution of planets, to probe the late stages of planetary
accretion (Schlichting and Sari, 2007).

A precise determination of the rotation rate can also help improve
our analysis of future direct detections of exoplanets, including
photometric, spectroscopic, and potentially polarimetric
observations (Gaidos and Williams, 2004; Tinetti et al, 2006;
Monta\~n\'es-Rodr\'iguez et al., 2005; Stam et al, 2006; Williams
and Gaidos, 2007). For practical viewing geometries, most of the
light scattered by an Earth-like planet comes from a small portion
of the planet, and contains information about weather patterns,
surface features, i.e. lands and oceans.  While even the most
ambitious space telescopes will not be able to spatially resolve the
surface of an extrasolar planet, the temporal variability contains
information about regional surface and/or atmospheric features,
possibly including localized biomarkers (Ford et al., 2003; Seager
et al, 2005; Monta\~n\'es-Rodr\'iguez et al., 2006). Determining the
planet rotation period is necessary in order to know the rotational
phase for a time series of observations. The precision with which
the rotation period can be measured determines the time span of
observations that can be coherently averaged.

We will see in this paper how the deviations from a periodic
photometric signal can help to identify active weather on an
exoplanet. This could prove a useful technique for recognizing
exoplanets that have weather systems with inhomogeneous cloud
patterns.

Finally, the observations of our solar system bodies suggest that
the presence of a planetary magnetic field, generated by dynamo
processes, is mainly a function of two parameters: its composition
(mass) and the rotation speed (Vallee, 1998; Russell, 2006). If the
planet mass is known, a fast rotation speed of the planet could
suggest the presence of a significant magnetic field. One must note
however that there will be a large list of caveats to this
possibility, given our current understanding of dynamos and
planetary evolution (Bushby and Mason, 2004; Grie{\ss}meier, 2007).

In this paper we have determined the changes in photometric albedo
that we would see if Earth was observed as an extrasolar planet.
First, we perform an accurate and realistic simulation of the flux
changes in reflected light from the planet's surface and atmosphere.
Second, we perform a periodicity analysis to determine under what
conditions the rotation rate can be determined.  Third, we explore
how the accuracy and precision of the measured rotation rate depend
on four variables: the temporal resolution of observations (i.e.,
exposure time), the total duration of observations, the
signal-to-noise ratio, and the viewing geometry. We also
discuss the role of clouds in altering the reflected light flux
from Earth, and how to detect them in an exoplanet's atmosphere.
Finally, we discuss the implications for the design of future space
missions to characterize extrasolar planets via direct detection.

\section{Methods}

\subsection{Planet Light Scattering Model}

The albedo of each surface element, $a$, depends on the surface
type, cloud and snow/ice cover and solar zenith angle. Further,
there is an anisotropic factor, $L$, that gives the angular
distribution of the reflected radiation and depends upon the
reflected zenith angle and azimuth. The anisotropy function, also
known as bidirectional reflectance function (BDRF), generally
depends on surface type, cloud cover, zenith angle and relative
azimuth of the Sun. $L$ is defined so that it is unity for a Lambert
surface (Pall\'e et al, 2003). In modeling the reflectance
properties, $a$ and $L$, of the Earth, we used scene models based on
the Earth Radiation Budget Experiment (ERBE) observations (Suttles
et al., 1988), defined as the mean over the broad shortwave interval
from 200 to 4000 $nm$. The parameters $a$ and $L$ are tabulated for
twelve model scenes.

The model of the Earth uses daily satellite observations of total
cloud amount at each surface location from the International
Satellite Cloud Climatology Project (ISCCP) as input (Rossow et al.,
1996). Four cloudiness levels (0-5\%,5-50\%, 50-95\% and 95-100\%),
are considered for each of the 12 different ERBE scenes. For the
snow/ice cover, we used simulations from the Canadian Center for
Climate Modeling and Analysis (CCCM II). The model has already been
validated by observations of Earthshine (Pall\'e et al, 2003).

Our model allows us to simulate the Earth's reflectance observed
from any viewing geometry.  For example, looking at the exoplanet
(our modeled Earth) always from the north pole or along the
ecliptic. In the context of observing extrasolar planets, this is
similar to fixing the orbital inclination of the orbit with respect
to the observation point. Thus, the Earth's reflectance in the
direction of $\beta$, where $\beta$ is defined as the angle between
the Sun-Earth and Earth-Observer vectors, can be expressed as

\begin{equation}
p_{e}f_{e}(\beta)={1\over\pi R_{e}^2} \int_{(\hat R\cdot\hat S,\hat
R\cdot\hat M)\ge 0} d^2R (\hat R\cdot\hat S)a(\hat R\cdot\hat M)L,
\label{Gamma}
\end{equation}
where $\hat R$ is the unit vector pointing from the center of the
Earth to a patch of Earth's surface, $\hat S$ is the unit vector
pointing from the Earth to the star, and $\hat M$ is the unit vector
pointing from the Earth toward the observer.  The integral is over
all of the Earth's surface elements for which the sun is above the
horizon (i.e., $\hat R\cdot\hat S$) and the surface element is
visible from the observer's perspective (i.e., $\hat R\cdot\hat M\ge
0$). Here $R_e$ is the radius of the Earth, $p_e$ is the geometrical
albedo of the Earth, and $f_e(\beta)$ is the Earth's phase function
(defined such that $f_e(0)=1$).

The total reflected flux in a given direction, $\beta$, can be
calculate using
\begin{equation}
F_{e}(\beta) = S \pi R_{e}^{2}  p_{e}f_{e}(\beta), \label{Fe}
\end{equation}
where $S$ is the solar flux at the top of the Earth's atmosphere
($1370 W/m^{2}$). There is a systematic variation of
$p_{e}f_{e}(\beta)$ throughout the Earth's orbital period (sidereal
year), and fluctuations of $p_{e}f_{e}(\beta)$ about its systematic
behavior are caused by varying terrestrial conditions, including
weather and seasons (Pall\'e et al, 2004).

\begin{figure*}[h]
\epsscale{1.0}  \plotone{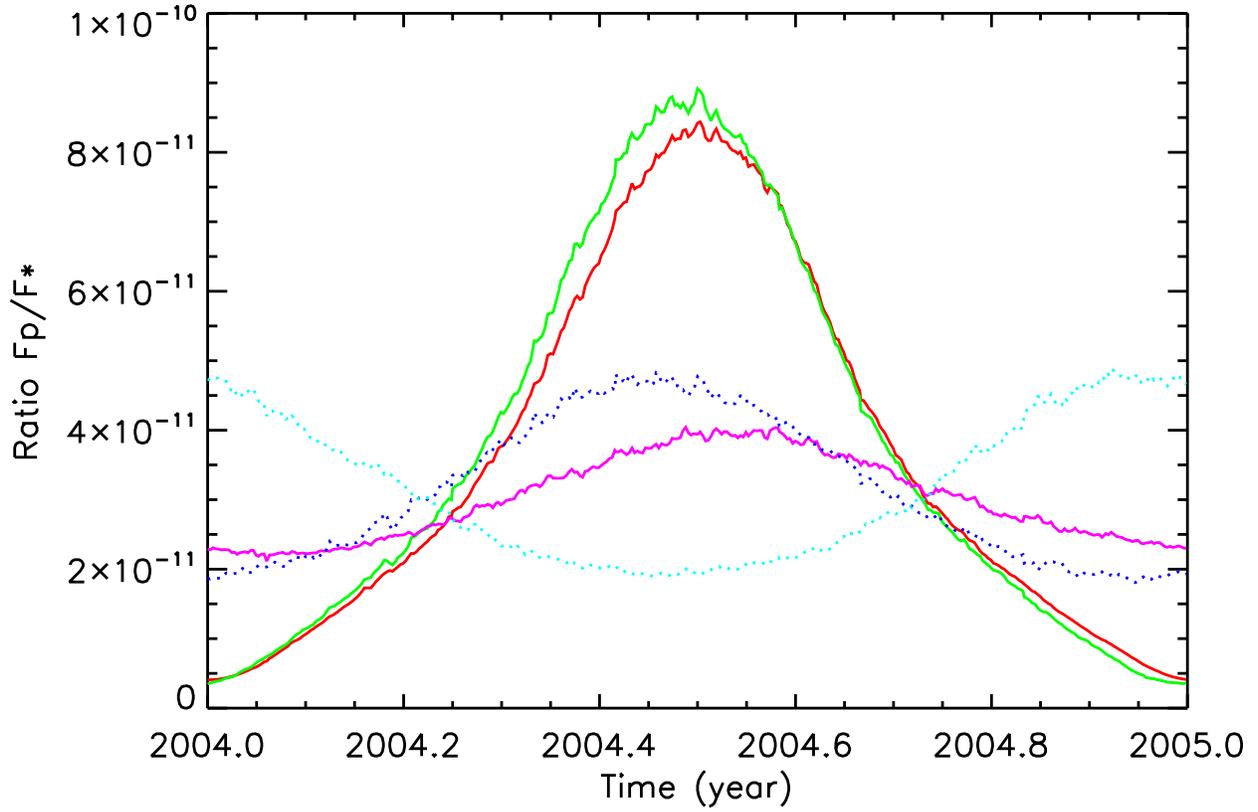} \caption{The yearly evolution of
relative flux of the Earth with respect to the Sun from five
different viewing geometries. The equatorial veiw is marked in red,
the primarily norther/southern hemisphere views are in green and
pink (respectively) and the north and south polar views are in dark
and light blue (respectively). } \label{fig3}
\end{figure*}

Comparing $F_e$ to the flux of sunlight for the same observer,
yields contrast ratios of order $10^{-10}$.  This presents the main
challenge in directly detecting an Earth-like planet. In comparison,
the amplitude of the diurnal cycle of the Earth observed in our
broadband ($200-4000 nm$) simulations (Figure~\ref{fig3}) is of the
order of $0.5 \times 10^{-11}$, but varies greatly depending on
wavelength.

At present, broadband coronagraphic experiments are able to reach
contrasts of $10^{-3}$ only (Mawet et al., 2006). However, advances
in the development of coronagraphs and deformable mirrors are
expected to enable such observations in the future.  For example,
Trauger and Traub (2007) have shown how contrast ratios of the order
of $1\times 10^{-11}$ can be achieved with coronagraphs in the
laboratory, using a laser beam at monochromatic visible wavelength.
In this paper, however, a wide bandwidth is considered, in order to
have enough photons in each observation. The use of a wide bandwidth
in coronagraphy will require a very good achromatization of the
coronagraph to achieve a high light rejection, working towards a
viable visible-wavelength direct imaging technique.

\subsection{Viewing Geometry}
\label{SecGeo}

In order to simulate the observations of the Earth as if it were a
distant planet, we must specify the viewing geometry of the
simulated observations. An observer that looks at the Sun-Earth
system from along the ecliptic will be looking at the Earth from a
nearly equatorial perspective. During a year the Earth will appear
to go through phases from a fully lit Earth to a fully dark Earth.
For this case, the Earth would pass inside of the Sun's glare twice
per year.  On the contrary an observer looking at the Sun-Earth
system from a direction perpendicular to the plane of the Earth's
orbit, would see only the northern (or southern) hemisphere of the
Earth. At any given time, approximately half of the Earth would be
illuminated and visible to the distant observer.

\begin{figure*}[h]
\epsscale{0.5} \plotone{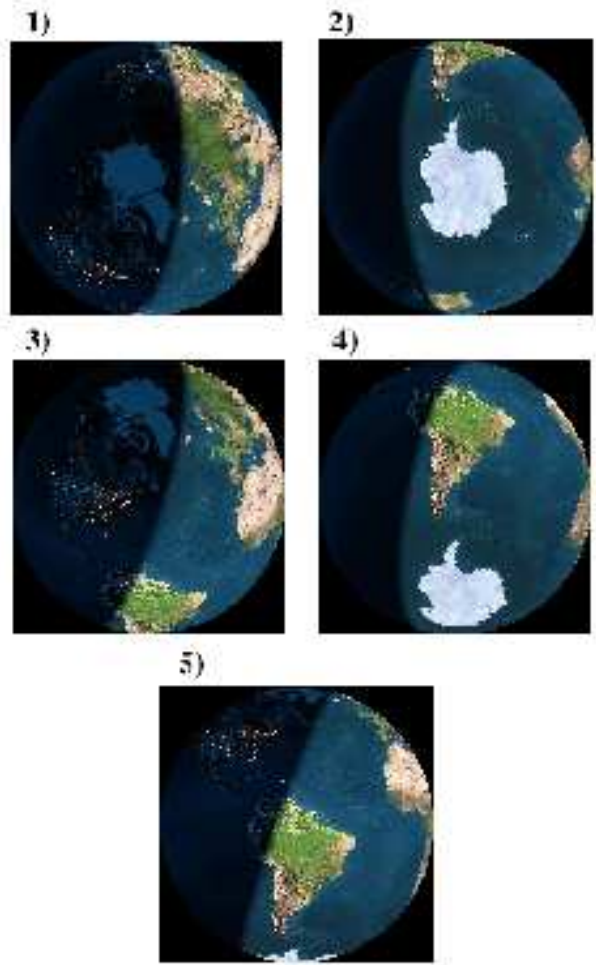} \caption{The Earth from space. The
several images shown the viewing geometry of Earth for the exact
same day and time (2003/11/19 at 10:00 UT) but from our five
different viewpoints: from $90^\circ$ above the ecliptic (north
polar view) (1), from $90^\circ$ below the ecliptic (south polar
view) (2), from +45$\circ$ north of the ecliptic (primarily northern
hemisphere in view) (3), from $-45^\circ$ below the ecliptic
(primarily southern hemisphere in view) (4), and from within the
ecliptic (5). Note how the scenery from the different viewpoints,
could well have been taken from different planets. } \label{fig2}
\end{figure*}

In order to determine the sensitivity of our results to viewing
angles, we have chosen five different viewing geometries of the
Earth which we will refer to as: the equatorial view, the
north/south polar view, and the (primarily) northern/southern
hemisphere view. Technically, we are choosing the inclination ($i$)
of the line of sight with respect to to the ecliptic plane:
0$^\circ$, $\pm45^\circ$, and $\pm90^\circ$. In order to visualize
the viewing geometries, Figure~\ref{fig2} shows the Earth for a
single date and time, as seen from each of the five viewing
perspectives that we consider.  The date corresponds to a day in
November, when the Earth would present a phase angle of
approximately $90^\circ$ (as seen from each of our viewpoints). Note
that the figure is misleading in the sense that clouds, which will
play a major role in the photometric albedo, are not represented.

The quantity $p_{e}f_{e}(\beta)$ is affected by three factors.
First, as the Earth's revolves around the Sun, $p_{e}f_{e}(\beta)$
will change due to a changing $\beta$. At the same time, due to the
Earth's rotation, the portion of the Earth's surface visible to the
observer will also change, leading to changes in the albedo diurnal
cycle. Finally, the large-scale cloud patterns will change from day
to day, adding short-term variability to the observations. In
Figure~\ref{fig3} the yearly evolution of the flux ratio between the
Earth and the Sun, taking into account these various effects, are
represented.

We generate photometric time series of the light scattered by the
planet toward an observer that include the effects of both the
planet's rotation and the planet's orbital motion (as well as
changing cloud and snow/ice cover).  While our simulated data is
centered on a phase angle of $\beta = 90^\circ$, the phase angle
deviates from $90^\circ$ due to the orbital motion (e.g., up to
$\simeq 28^{\circ}$ for an eight week time series with the
equatorial viewing geometry).

\subsection{Observational Considerations}

Several considerations need to be taken into account before we can
realistically analyze our simulations in terms of the Earth as an
exoplanet. A space telescope intending to the search for exoplanets
will have a long list of target stars to observe during the planned
mission life time (of order a few years).  If a small number of
remarkable Earth-like planet candidates are identified, then
multiple months of observations time could be devoted to
characterizing individual targets.  On the other hand, if many
Earth-like planet candidates are found, then the amount of observing
time available for follow-up observations of most targets could be
much more limited.  Therefore, we have considered simulated
observational data sets spanning 2, 4, and 8 weeks. Similarly, we
have simulated observations made with several temporal resolutions
(exposures times), ranging from 0.1 to 10 hours. Finally, we have
added Poisson noise to the data, to simulate signal-to-noise (S/N)
ratios, for each exposure time, ranging from 3 to 1000. While such a
large S/N is unrealistic for a first generation TPF-C mission, these
calculations are relevant for determining if very high-precision
rotation measurements are possible or if the stochastic nature of
clouds results in a limit on the precision of rotation periods that
is independent of the S/N.

The orbital position of the planet will also limit our observing
capabilities. Ideally one would like to observe the planet at full
phase when its brightness, as compared to that of the parent star,
is larger. However, observations at these phase angles are nearly
impossible due to the small angular distance between the planet and
the star. In this work, we focus on observations made near a phase
angle of $90^\circ$, when the planet-star separation is near its
maximum. The best case scenario for measuring the rotation period of
a planet occurs for an orbital plane nearly perpendicular to the
line of sight, so that the planet remains at a phase angle of nearly
$90^{o}$ (maximum angular separation) for it's entire orbit orbit.

\section{Results}

\begin{figure*}[h]
\epsscale{1.0} \plotone{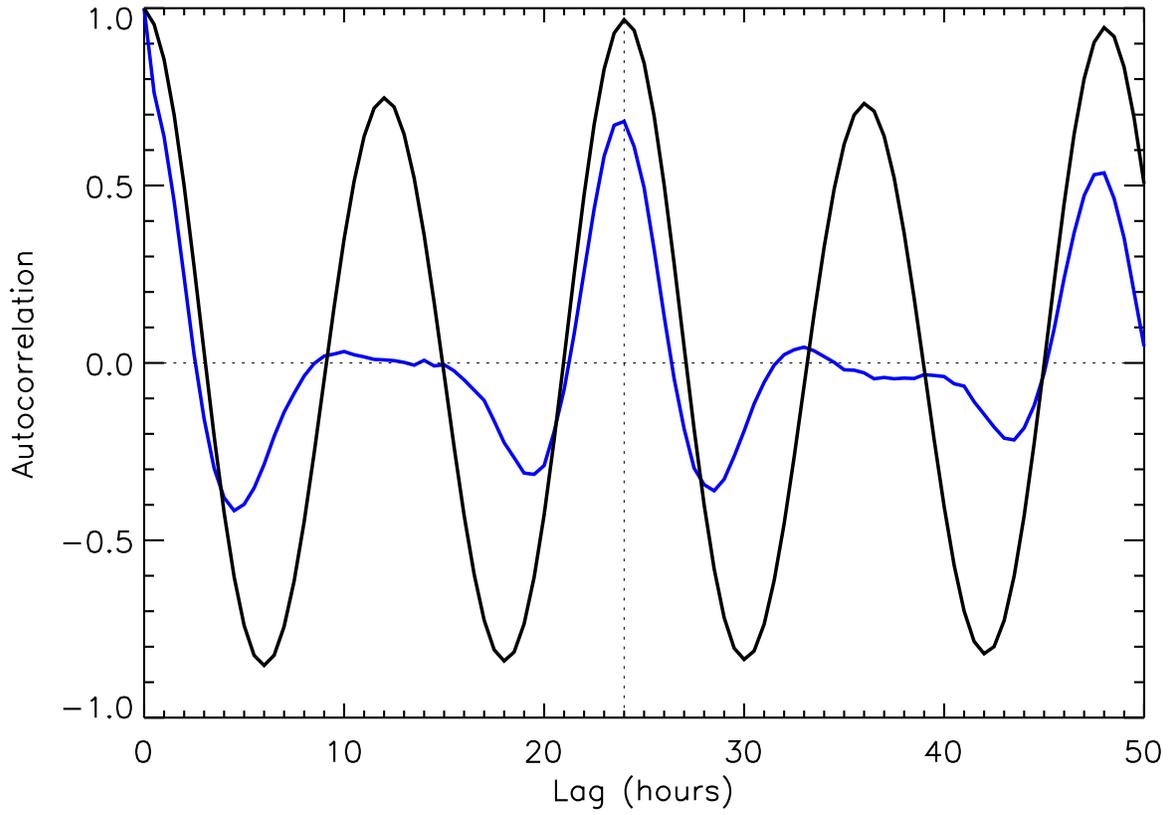} \caption{Autocorrelation function of
the scattered light by the real Earth (blue) and a cloud-free Earth
(black). An 8-week time series with S/N ratio of 40 and 0.1 hour
observing cadence has been chosen, using cloud data from 1985.}
\label{fig6}
\end{figure*}

We simulate several time series of the Earth's scattered light
towards a hypothetical observer.  For each time series, we perform
an autocorrelation analysis.  For example, in Figure~\ref{fig6}, the
black curve shows the autocorrelation as a function of the time lag
based on a simulated data series for an Earth without any cloud
cover.  We assume the $i=90^\circ$ viewing geometry described in
\S\ref{SecGeo} and observations with a signal-to-noise ratio of 40
and 0.1 hour temporal resolution. Such assumptions are clearly
optimistic, but not completely unreasonable. A 8m x 3.5m TPF-C
mission could make such a high precision measurement for an
Earth-like planet (i.e., 25 magnitudes fainter than the host star)
with a 24 hour rotation period in the Alpha Centauri system (based
on a 400 nm bandpass centered on 650nm, an extrasolar zodiacal light
equal to that of the solar system, and the algorithm and other
assumptions from ``case A'' of a TPF-C mission described in Brown
(2005).  Therefore, if terrestrial planets are ubiquitous, then even
a first generation TPF-C mission may be able to determine the
rotation periods of terrestrial planets with high precision.

In \S3.2 we will show that the rotation period can be measured with
moderate precision using only a signal-to-noise ratio of $\sim$20
every $\sim$16th of a rotation period.  An 8mx3.5m TPF-C mission
could achieve such photometric precision for stars brighter than
V=4.  In \S5, we will further discuss the capabilities of such a
TPF-C mission, as well as missions of alternative sizes. Based on
the TPF Target List Database (v1; http://sco.stsci.edu; see also
Turnbull and Tarter, 2003), we find that 29 such main sequence K-A
stars that have accurate parallax, B-V color, no companion stars
within $10 arcsec$, and show no indications of variability.
Eliminating A stars reduces the number of such targets to 15.  Note
that this is more than the 14 and 5 target stars included in the TPF
Target List Database `extended'' (including A stars) and ``core''
(excluding A stars) lists that apply several additional cuts based
on a notion of habitability (e.g., eliminating young stars that may
be too young for significant biological alteration of the
atmosphere).

\subsection{Measuring the Rotation Period}

By definition, the maximum autocorrelation equals unity at zero lag.
The next greatest autocorrelation occurs at 24 hours, very near the
true rotation period of the Earth.  In this case, we find that the
amplitude of the autocorrelation is very similar at integer
multiples of the Earth's rotation period, since the only changes are
due to the slow variations of the viewing geometry and phase angle
($\beta$) resulting from the orbital motion of the Earth.

For a cloudless Earth, we find that there is a second series of
local maxima in the autocorrelation function near twelve hours. This
is not due to a fundamental property of the autocorrelation analysis
(e.g., the blue curve for the cloudy Earth has no significant
amplitude at 12 hours), but rather is due to the distribution of
continents and oceans on the Earth.  For this data set, the
difference in the amplitude between the local maxima at 12 hours and
24 hours would indicate that the peak at 24 hours corresponds to the
rotation period.  However, the possibility of the continental
distribution leading to a significant autocorrelation at alternative
lags could complicate efforts to identify the rotation period.

We now consider a cloudy Earth using Earth's actual cloud cover
randomly selected for eight weeks in 1985. The blue curve in
Figure~\ref{fig6} shows the results of an autocorrelation analysis
similar to the one for the cloud-free Earth, assuming the same
viewing geometry and observational parameters as above.  Aside from
the maximum at zero lag, the maximum autocorrelation occurs at 24
hours, very near the true rotation period of the Earth. The
additional local maxima of the autocorrelation that occur at integer
multiples of 24 hours are due to the viewing geometry repeating
after multiple rotations of the Earth. In this case, the the
autocorrelation decreases at larger multiples of the rotation
period, since the variations in the cloud patterns are typically
greater on these longer time scales.

\subsection{Accuracy and Limits in the Measurements}

Here, we explore how the precision of the measured rotation period
depends on various observational parameters, such as the
signal-to-noise ratio, the temporal resolution of observations, the
total duration of the observational campaign, and the viewing
geometry.

\begin{figure*}[h]
\epsscale{0.6} \plotone{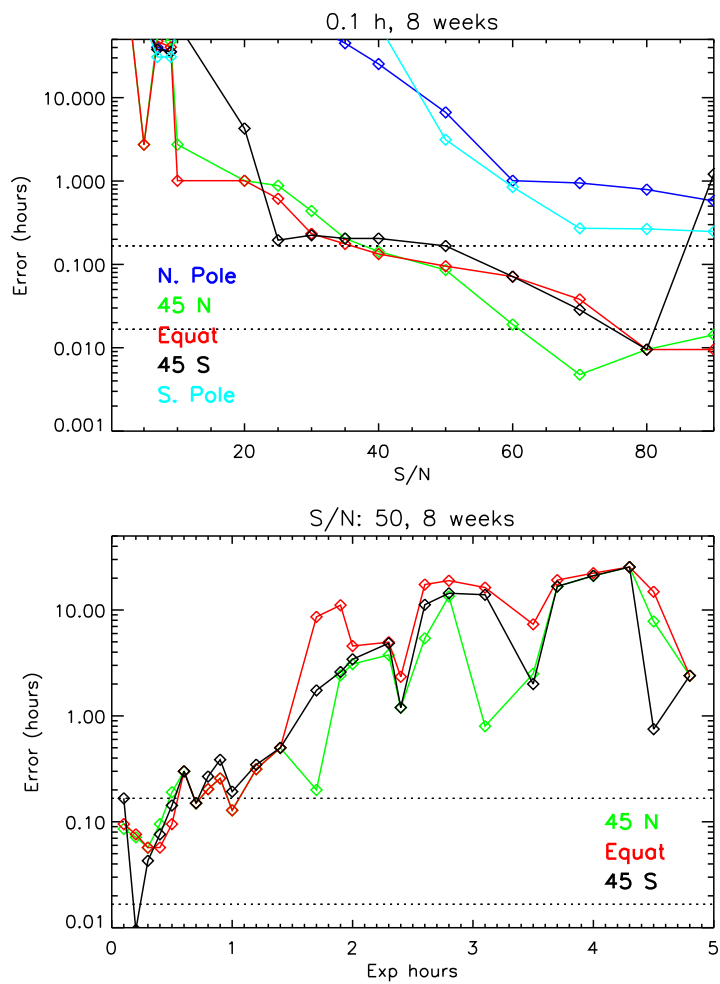} \caption{Top panel: The plot
represents the error that one would get in estimating the Earth's
rotation rate from the globally integrated photometric light curve.
Each point is the error of the averaged rotational period found for
21 years with different (real) cloud patterns for the same
geometries. The five different colors indicate five different
viewing angles (i.e equator means the observer is looking at the
Sun-Earth system from the ecliptic plane, the North pole indicates
the observer is looking at the Sun-Earth system from $90^\circ$
above the ecliptic). All calculations are given for a $90^\circ$
phase angle in the orbit (i.e. one would see a quarter of the
Earth's surface illuminated). In the plot, the top broken line
represents an accuracy in determining the rotational period of
$10~minutes$, and the lower one of $1~minute$. Lower panel: Same as
in the top panel, but this time the S/N is fixed and the exposure
time is allowed to vary. As in the top panel, an object follow up of
two months (8 weeks) is considered.} \label{fig7}
\end{figure*}

In Figure~\ref{fig7} (top), we show the mean absolute value of the
difference between the actual and the derived rotation period of the
Earth based on 21 data sets, each for a different year (Global cloud
coverage measurements from ISCCP satellite observations are only
available over the period 1984-2005, i.e. 21 years). Here we assume
8 weeks of observations with a temporal resolution of 6 minutes.
Each curve corresponds to a different viewing geometry. For the
equatorial and primarily northern/southern hemisphere views, we
conclude that a S/N ratio of 10-20 is necessary to determine the
rotational period with an error of about 1 hour (5\% of the 24-hour
rotation period). With a S/N ratio of about 30, we find a precision
in the rotation determination of approximately 10 minutes (0.7\%).
On the other hand, the determination of the rotational period from a
polar perspectives has a larger error. Even with increasing S/N, the
rotational period that one obtains from a polar perspectives does
not always converge to 24 hours but to a shorter periodicity (see
\S\ref{SecRot} for further discussion).

In Figure~\ref{fig7} (bottom), we show the mean absolute value of
the difference between the actual and derived rotation period of the
Earth, but as a function of the temporal resolution of the
photometric observations.  We assume a fixed signal-to-noise of 50
and an 8 week observing campaign.  It is clear from the figure that
a temporal sampling no larger than 1.5 hours (6\% of the period) is
desirable, if we want to derive the rotational period with
precision. Again we find very different results for the polar
viewing geometries than for the rest of viewing geometries (not
shown). For the polar geometries, the integration time is not the
key factor in determining the rotation.

For the equatorial view and primarily northern/southern hemisphere
views, the rotation period can be determined accurately, provided
that the exposure time is shorter than 1.5 hours. For exposure times
larger than about 1.5 hours, the periodicity near 12 hours might be
mistaken for the true rotation period.

For a general planet, we expect that the temporal resolution needed
will scale with the planet's rotation rate.  E.g., a similar planet
with a rotation period of 8 hours, would require that the exposure
times be reduced by a factor of three to achieve a similar precision
in the determination of the rotation period.

\subsection{Autocorrelation vs Fourier}

\begin{figure*}[h]
\epsscale{1.0}\plotone{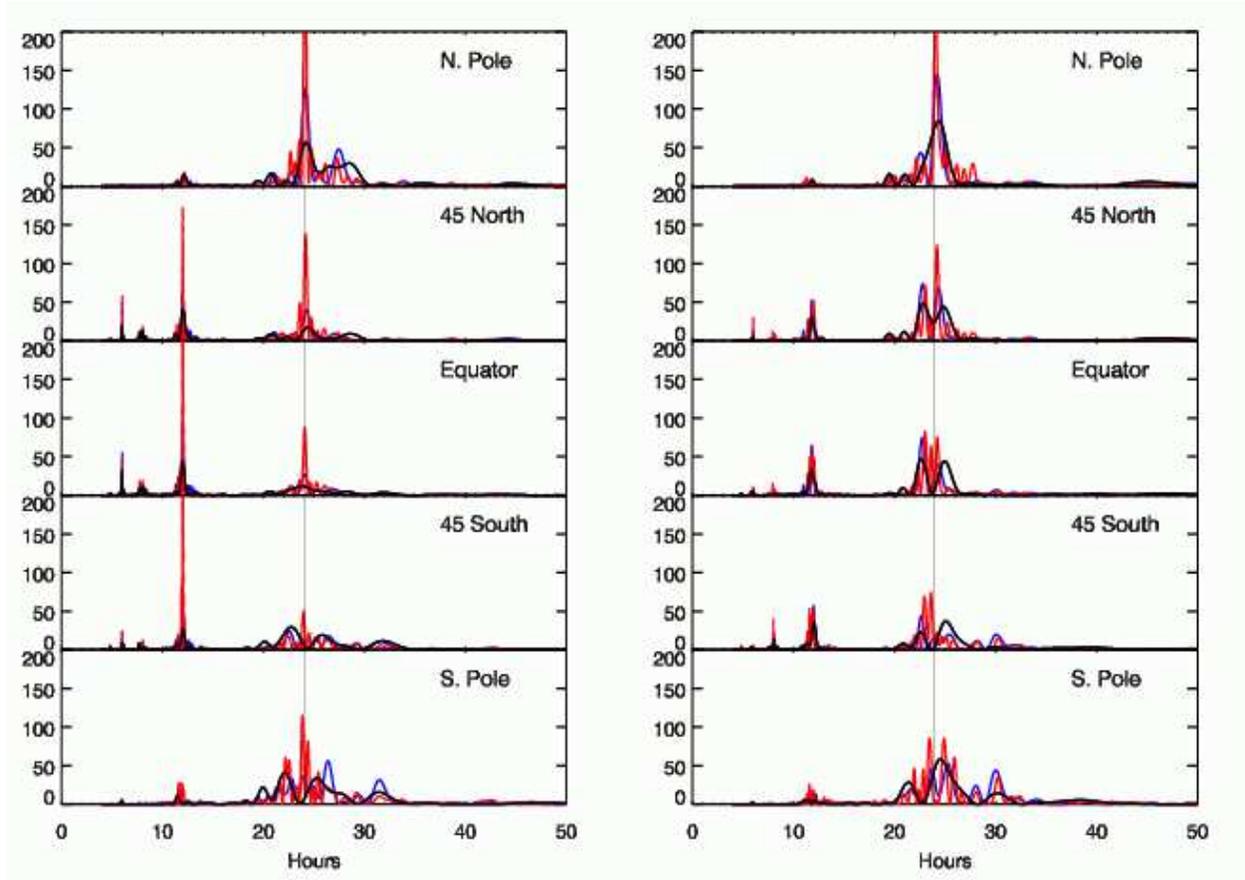} \caption{Periodogram analysis of the
Earth's $p_{e}f_{e}(\beta)$ times series as seen from five different
viewpoints, at phase angle $90^{o}$. From top to bottom, the five
viewpoints are: the north polar view (a), primarily northern
hemisphere view (b), the equatorial view (c), primarily southern
hemisphere view (d), and the south polar view (e). The right column
represents the periodograms for the year 2000, while the left column
represents the periodograms for year 2004. The geometry is exactly
the same for the two years, only the clouds have changed. In all
panels, the periodogram is shown for data lasting for a period of 2
(black line), 4 (blue line) and 8 weeks (red line) around phase
$90^{o}$. In all panels a thin vertical line indicates the ``real"
24-hour periodicity. } \label{fig5}
\end{figure*}

Here, we explore the outcome of performing a periodicity analysis to
our simulated photometric time series, using a Fourier-based
technique, the classical periodogram. In Figure~\ref{fig5} we have
plotted the periodogram of the time series resulting from
simulations of the real (cloudy) Earth, as viewed from the five
different viewpoints and in two different years. For each case, the
periodogram is calculated using time series spanning 2, 4 and 8
weeks of observations.

According to the periodogram analyses, the 24-hour periodicity is
not always the strongest, and it is missing altogether for some of
the series (depending on the specific cloud data). The observations
from the nearly equatorial perspective seem to be the most affected
by the Earth's particular continental distribution, as there are
strong peaks at 12-hours. If the distribution of continents on our
planet were different (as it has been in the past), then the derived
periodicities would also be different.  The viewing geometry also
plays a role. For example, our southern pole viewing geometry
results in the Earth appearing to have a single continent in the
center surrounded by ocean.

We compare the accuracy and precision of two types of periodicity
analysis: the autocorrelation function and Fourier analysis.  In
Table~\ref{tab1}, we show the frequency with which each type of
analysis results in a determination of the rotation period near the
true rotation period ($24\pm \Delta$ hours), half the rotation
period ($12\pm \Delta$ hours), or other alternative values. In each
case $\pm \Delta$ is taken as the exposure time (or sampling
resolution). In other words, a $95\%$ value in the 24-hour
periodicity for the autocorrelation method, means that for 20 of the
21 available years (cloud configurations) the primary periodicity
retrieved by the autocorrelation method is $24\pm \Delta$. In
Table~\ref{tab2}, we present the the same quantities as in
Table~\ref{tab1}, but this time for an Earth completely free of
clouds, so that the detected periodicities are due to surface albedo
variations only.  The Fourier analysis often results in the largest
peak near 12 hours (see Fig.\ \ref{fig5}), particularly for viewing
geometries with orbital inclinations of $45^\circ$, $90^\circ$, and
$135^\circ$. Our autocorrelation analysis never makes this error.
Thus, we conclude that the autocorrelation function provides a more
robust and more accurate tool for characterizing the rotation period
of a planet using photometric time series data.

\section{The Effect of Clouds}

Clouds are common on solar system planets, and even satellites with
dense atmospheres.  Clouds are also inferred from observations of
free-floating substellar mass objects (Ackerman and Marley, 2001).
Hence, cloudiness appears to be a very common phenomenon.

Clouds on Earth are continuously forming and disappearing, covering
an average of about $60\%$ of the Earth's surface.
This feature is unique in the solar system to Earth: Some solar
system planets are completely covered by clouds, while others have very few.
Only the Earth has large-scale cloud patterns that partially cover the
planet and change on timescales comparable to the rotational
period.  This is because the temperature and pressure on the Earth's
surface allow for water to change phase with relative ease from
solid to liquid to gas.

In principle, weather patterns and/or the orbital motion of the
Earth could pose a fundamental limitation that prevents an accurate
determination of the Earth's rotation period from the scattered
light. Since the scattered light is dominated by
clouds, it might be impossible to determine the rotation period if
the weather patterns were completely random.  Alternatively, even if
the atmospheric patterns were stable over many rotation periods,
observational determinations of the rotation period might not
correspond to the rotation period of the planet's surface, if the
atmosphere were rotating at a very different rate (e.g., Venus).

\begin{figure*}
\epsscale{1.0} \plotone{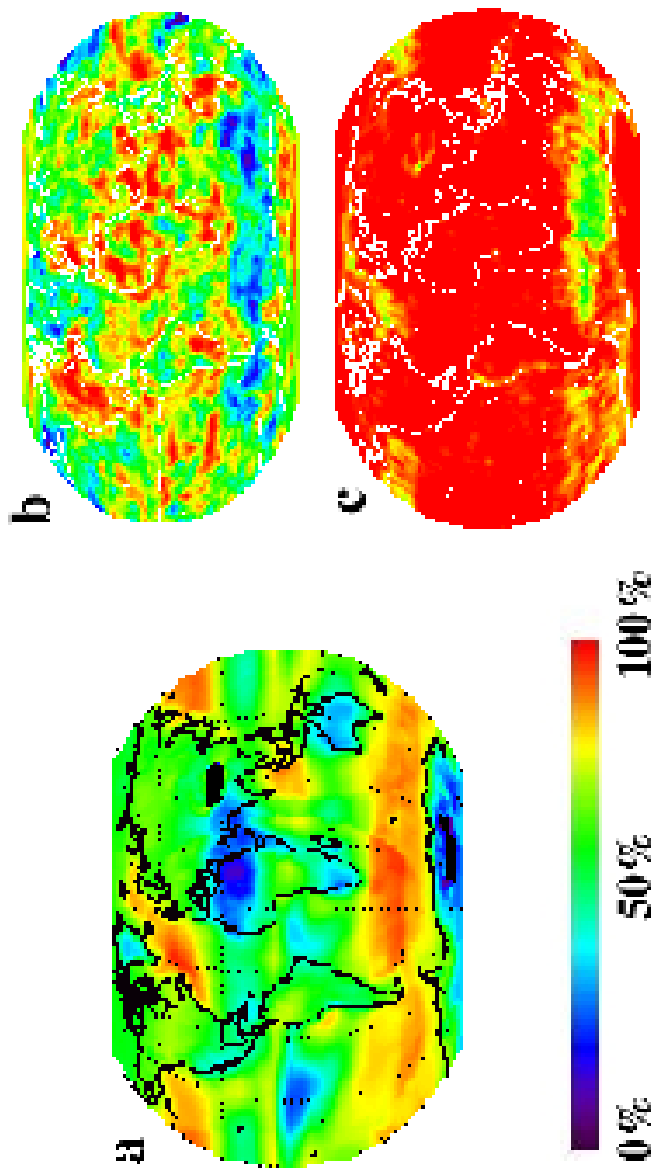} \caption{Large-scale cloud
variability during the year 2004. In panel (a) the 2004 yearly mean
cloud amount, expressed in percentage coverage, is shown. In panels
(b) and (c) cloud coverage variability (ranging also from 0 to
100\%), is illustrated over a period of 2 weeks and 1 year,
respectively. Over the course of 2 weeks, the presence of clouds at
a given location is highly correlated. Note how the cloud
variability is larger at weekly time scales in the tropical and
mid-latitude regions rather than at high latitudes. Over the course
of a whole year the variability is closer to 100\% over the whole
planet (i.e., at each point of the Earth there is at least a
completely clear and a completely overcast day per year). One
exception to that occurs at the latitude band near $-60^{o}$, an
area with heavy cloud cover, where the variability is smaller, i.e.,
the stability of clouds is larger. } \label{fig9}
\end{figure*}

In fact, we find that scattered light observations of the Earth
could accurately identify the rotation period of the Earth's
surface. This is because large-scale time averaged cloud patterns
are tied to the surface features of Earth, such as continents and
ocean currents. This relatively fixed nature of clouds (illustrated
in Figure~\ref{fig9}) is the key point that would allow Earth's
rotation period to be determined from afar.

Figure~\ref{fig9} shows the averaged distribution of clouds over the
Earth's surface for the year 2004. The figure also shows the
variability in the cloud cover during a period of two weeks and over
the whole year 2004. The lifetime of large-scale cloud systems on
Earth is typically of about 1-2 weeks (roughly 10 times the
rotational period). In the latitude band around $60^{o}$ south,
there is a large stability produced by the vast, uninterrupted
oceanic areas.

\subsection{The Folded Light Curves}

\begin{figure*}
\epsscale{1.0} \plotone{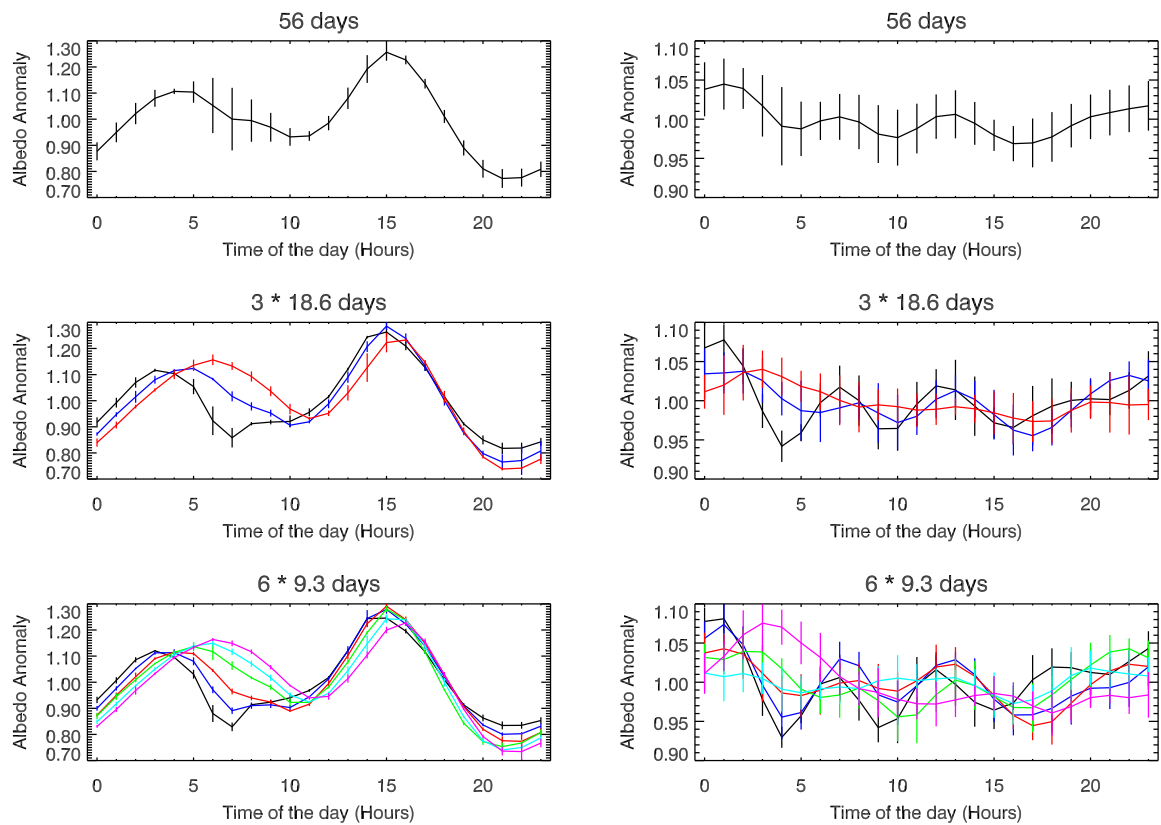} \caption{Light curves of the Earth
observed from the ecliptic plane at phase $90^{o}$ (half phase).
Left column are the light curves of a cloud-free Earth and right
columns are the light curves for the real Earth. The Y-scale in the
right and left panels is different because of the more muted
variability in the albedo introduced in the real Earth by clouds.
Fifty six days (two months) of continuous observations are divided
from top to bottom in 1, 3, and 6 sub-series, and folded over the
24-hour rotational period of the Earth for analysis. Note the
contrast between the uniformity of the light curves of an ideal
(cloudless) Earth and the real Earth light curves. Also note how the
change in the shape of the light curve is smooth (ordered in time)
from one series to the next in the case of a cloudless Earth, but it
is random for the real Earth. The size of the error bars are the
standard deviation of the mean.} \label{fig8}
\end{figure*}

In Figure~\ref{fig8}, we show the folded light curve of the Earth in
terms of the albedo anomaly, both with and without clouds. Albedo
anomaly is defined as the standard deviation (rms) from the mean
value over the entire 8-week dataset (e.g., an anomaly value of 0.7
means that the albedo in 30\% lower than the mean). Here we assume
an exposure time of 1 hour and S/N=30. The real Earth presents a
much more muted light curve due to the smoothing effect of clouds,
but the overall albedo is higher. Note that the $Y$ scale in the
figure are anomalies and not the absolute albedo values.

In the top panels of Figure~\ref{fig8} data from 8 weeks of
continuous observations are folded into a single light curve. In the
middle and lower panels, this 8 week period is subdivided and
plotted in 3 and 6 periods of 18.6 and 9.3 days, respectively. For a
cloudless Earth (left panels), the error in the albedo anomaly at a
given phase decreases as shorter durations are taken, because
changes in phase and illuminated area decrease.

On the contrary, for the real cloudy Earth (right panels), as the
data is subdivided in smaller integration periods, the size of the
error bar in the albedo anomalies does not decrease, because of the
random influence of clouds at short time scales. In the lower right
panel of Figure~\ref{fig8}, the light curves of consecutive 9-day
integration periods vary arbitrarily in shape from one to the next.

Thus, the variability in the averaged light curve is primarily the
result of short-term variability in the cloud cover, a fact that can
be exploited in future exoplanet observations. Once the rotational
period has been determined, one can measure the average light curve
of an exoplanet, and the excess scatter for different consecutive
periods. If the excess scattering does not decrease at short time
periods, and the changes are not smooth in time, such an analysis
can indicate the presence of clouds in its atmosphere. However,
distinguishing the changes in the exoplanets light curve from the
observational noise will require very stringent S/N ratios.
Fortunately, there might be a better way to probe for cloudiness in
an exoplanet's atmosphere that we discuss in the following section.

\subsection{Real and Apparent Rotational Period}
\label{SecRot}

\begin{figure*} 
\epsscale{1.0} \plotone{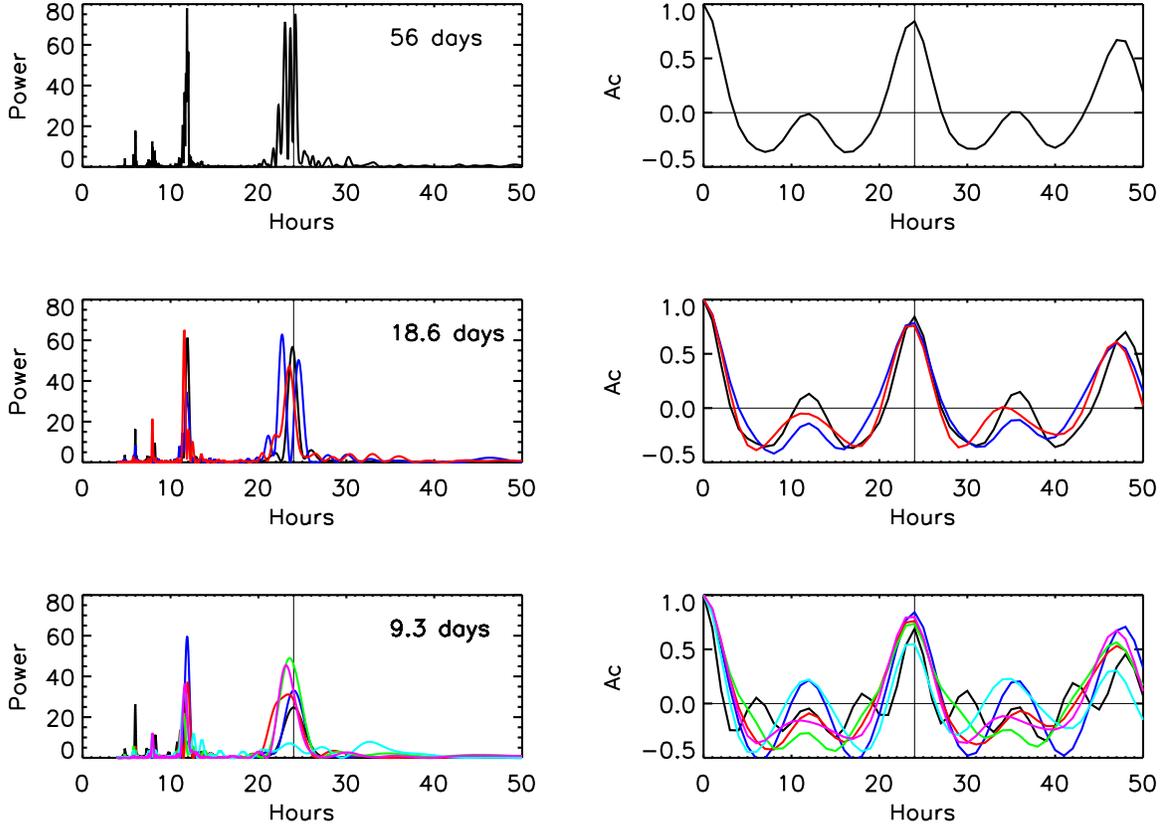} \caption{Left: Periodogram analysis
of the Earth's $p_{e}f_{e}(\beta)$ equatorial time series. Right:
Autocorrelation function of the same time series. For this figure we
select 8 weeks of data (56 days) and calculate the periodograms and
autocorrelation functions (top panels). S/N ratio are set here to 50
for clarity purposes. Then we subdivide these data in three (middle
panels) and six (bottom panels) equally-long time series and we
again calculate the separate periodograms and autocorrelations. In
the figures, different colors indicate different data subperiods.
Note the appreciable decrease in the retrieved rotation rate for
some of the time series in the bottom panels, detectable with both
autocorrelation and Fourier analysis.} \label{fig10}
\end{figure*}

For extrasolar planet observations, a long time series could be
subdivided into several subsets.  Each can be analyzed for
significant periodicities as in Figure~\ref{fig10}.  The data
spanning for 8 weeks is subdivided in several equal-length
subperiods (e.g., six periods each of about 9 days) and analyzed
independently, so that the changes in $\beta$ and illumination area
are minimized.  In this case, several peaks appear in the Fourier
periodograms and autocorrelation functions near 12 and 24-hours. The
autocorrelation analysis show much greater correlation near 24
hours. For our Earth model with clouds, the best-fit rotation period
shifts slightly to shorter periods. The shifts in the best-fit
periodicity from the true periodicity are completely absent when
considering an Earth model free of clouds for the same dates and
times, even when including added noise. Therefore, we conclude that
they are produced by variable cloud cover.

The shifts are introduced by the large-scale wind and cloud patterns
(Houghton, 2002). Since  clouds are displaced toward the west (in
the same direction of the Earth's rotation) by the equatorial trade
winds (and to a minor extent by the polar easterlies) the apparent
rotational period should be shorter than the rotation period of the
surface. On the other hand, when clouds are moved toward the east
(in the opposite direction of the Earth's rotation) by the westerly
winds at mid-latitudes, the apparent rotational period should be
longer than the rotation period of the surface.

In principle, both longer and shorter periodicities could be present
in the periodograms, depending on the particular weather patterns.
In our models however, we often find shorter apparent rotation
rates, but not longer. The explanation probably lies in the
different mechanisms of cloud formation on Earth. In the tropical
regions most of the clouds develop through deep convection. This
deep convective clouds have a very active cycle and a short
lifetime, in other words, these cloud systems do not travel far. At
mid-latitudes however, deep convection does not occur, and large
weather and cloud systems remain stable (and moving) for weeks (Xie,
2004).

Thus, both observing changes (anomalies) in the apparent rotational
period and the amount of scatter about the phase-averaged light
curve, one can recognize variable cloud cover and distinguish it
from the presence of strong surface inhomogeneities, and the
presence of a cloud layer. Thus, photometric observations could be
used to infer the presence of a `variable' surface (i.e. clouds),
even in the absence of spectroscopic data.  This would strongly
suggest the presence of liquid water on the planet's surface and/or
in the planet's atmosphere, especially if the mean temperature of
the planet were also determined. This could be an early step in
selecting the most desirable targets for more intensive follow-up
and/or observations with future more advance missions with more
powerful spectroscopic capabilities.

\section{Implications for Future Missions}

Finally, we consider the implication for future space missions. We
have shown that the integrated scattered light from the Earth
contains enough information to determine Earth's rotation period.
However, realistic space missions will likely be photon-starved.
Here, we address whether precise measurements of the rotation period
might be practical with next-generation observatories. First, we
will ask for what mission specifications and target stars would it
be possible to measure the rotation period of an Earth-close to
$\sim2\%$ precision. This choice is based on our simulated analysis
of the Earth's light curve that show the rotation period can be
determined to an average of $\sim2\%$ from data spanning 56 days
with a signal-to-noise ratios of $\sim20$ or greater and an
integration times no longer than $\sim$1.4 hours. Our simulations
reveal a significant decrease in the precision of the measured
rotation period for lower signal-to-noise ratios or longer
integration times. Moreover, for extrasolar Earth-like planets,
other parameters, such as the viewing geometry or the continental
distribution, will play a major role in wether we will be able to
measure the rotational period and with which level of accuracy.
Future research should investigate whether certain wavelengths or
combinations of observations can provide more robust measurement of
rotation periods.

Proposed missions such as the Terrestrial Planet Finder coronagraph
(TPF-C), Darwin, or SEE-COAST, are still in the planning stages and
final specifications are not yet available. For the sake of
concreteness, we follow Brown (2005) and consider a TPF-C mission
with an 8m$\times$3.5m primary mirror observing an Earth-like planet
which is 25 magnitudes fainter than the host star (their ``case
A''), except that we assume an extrasolar zodiacal light that is
equal to that of the solar system. We find that a signal-to-noise
ratio of 20 can be obtained within 1.4 hours (for a 110nm bandwidth)
for host stars of magnitude $\sim$3.8 or brighter. Therefore, we
estimate that there are $\sim11$ such stars included in the possible
TPF-C target list of Brown (2005) around which an Earth clone's
rotation period could be measured to $\sim$2\%.  If we were to scale
up the primary mirror of TPF-C by a factor of two (16m$\times$7m),
then the limiting magnitude increases to V$\sim$4.4, and there are
$\sim35$ stars in the sample target list of Brown (2005) for which
an Earth clone's rotation period could be measured to $\sim$2\%.  If
a bandpass of $\sim400$nm were practical, then the limiting host
star magnitude might increase by roughly one magnitude, making it
possible to measure rotation periods for Earth-clones around
$\sim35$ or $\sim90$ stars, for the two mission scenarios.  We
caution that these last two figures are very approximate, since the
expressions of Brown (2005) break down for large fractional
bandpasses.

\begin{figure*} 
\epsscale{1.0} \plotone{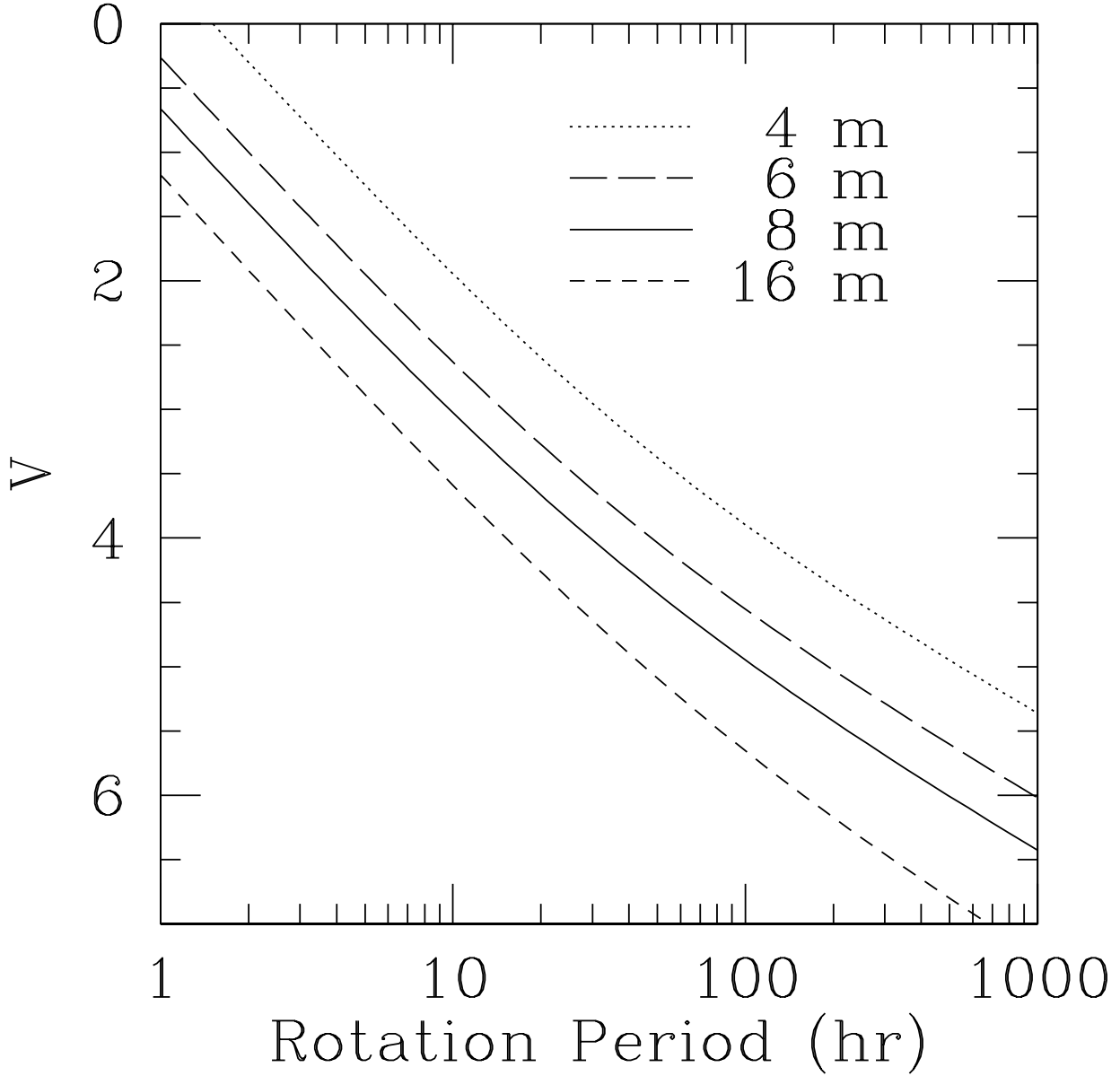} \caption{Here we show the threshold
host star magnitude and planet rotation period for which a
signal-to-noise of $\sim20$ or greater can be obtained for each
integration of $\sim$1.4 $(P_{\rm rot}/24)$.  Along this curve, our
simulations suggest that a times series spanning 56$\times P_{\rm
rot}$ would typically result in measuring the rotation period to
$\sim2\%$ for an Earth-like planet.  Higher precision measurements
of the rotation period would be obtained for $V$ and $P_{\rm rot}$
to the upper right of the curves.  The solid curve assumes mission
specifications similar to ``case A'' of Brown (2005), an Earth-like
planet that is 25 magnitudes fainter than the host star, and an
exozodi comparable to that of the solar system. The other curves
assume similar mission specifications, but scale the major axis of
the primary mirror to 4m (dotted), 6m (long dashed), or 16m (short
dashed) and hold the axis ratio constant.} \label{tpffig}
\end{figure*}

It would be somewhat easier to achieve the needed signal-to-noise
ratios for a planet that rotates more slowly than the Earth.  If we
were to ignore the effects of the planet revolving around the host
star, then our results could be scaled to apply to an Earth-like
planet with a rotation period of $P_{\rm rot}$.  For such a planet,
the threshold for achieving a rotation period precision of $\sim2\%$
would require achieving signal-to-noise ratio of 20 with integration
times of no more than $\sim$1.4 $(P_{\rm rot}/24)$. In
Figure~\ref{tpffig}, we show this threshold as a function of the
rotation period and the V magnitude of the host star.  The different
line styles indicate the assumed major axis of the primary mirror
(assuming the aspect ratio is held fixed at 16/7).  For Earth-like
planets above and to the right of these contours, we estimate that
the rotation period could typically be measured to $\sim2\%$ or
better.

The above estimates assume that the cloud patterns on the Earth
would not be affected by the alternative rotation period.  Further,
the above estimates also assume that the duration of the time series
scales with the rotation period of the planet.  A single continuous
time series would be impossible for a planetary system viewed nearly
edge-on, since the planet would periodically pass inside the glare
of the star (or inner working angle of the coronagraph).  Further
research is needed to determine how well the rotation period could
be measured by combining multiple shorter photometric time series,
and which are the most suitable spectral ranges. For a planetary
system with an orbital plane nearly in the plane of the sky, it
would be possible to obtain photometric time series spanning
$56\times P_{\rm rot}$, even for slowly rotating planets. Depending
on what other planets have been found, it might or might not be
practical to devote so much mission time to a single planetary
system.  We also caution that for planets with extremely slow
rotation periods that approach the orbital period (e.g., Venus), our
assumed scaling may break down due to seasonal effects and the large
changes in the viewing geometry.

\section{Conclusions}

Exoplanets are expected to deviate widely in their physical
characteristics and not all exoplanets will have photometric
periodicities. Some planets, such as Venus, are 100\% cloud covered
and show no significant photometric variability with time. A
variable photometric data set with no autocorrelation signal may be
indicative of slow rotation or chaotic weather.

On Earth, the presence of continents and ocean currents results in
relatively stable global cloud patterns, despite large variability
on short time and length scales. Here we have shown that, despite
Earth's dynamic weather patterns, the light scattered by the Earth
to a hypothetical distant observer as a function of time contains
sufficient information to measure Earth's rotation period to within
a minute, on the most favorable cases. The accuracy in the
rotational period determination is a function of the viewing
geometry, S/N ratio, temporal sampling and the duration of our
simulated time series. The rotation period could be directly
compared to numerical simulations of planetary formation, to probe
the late stages of planetary accretion.

According to our calculations, the duration of the observations is
comparable to the integration times needed for spectroscopic
observations to search for multiple atmospheric biomarkers (Traub et
al, 2006). Thus, we recommend that a photometric time series
spanning weeks to months be carried out simultaneously with planet
spectral characterization, via ``spectrophotometry''. Photon
counting CCDs have no read noise and are being adopted in mission
concept studies for TPF-C and related missions (Woodgate et al,
2006). Such photon counting CCDs tag photon arrival at different
wavelengths, and allows later binning in different ways.
Observations of an exoplanet spanning several weeks could be binned
over the entire observational period to retrieve a low-resolution
spectra and characterize its atmospheric composition. Additionally,
the data could also be binned in shorter time periods over all
wavelengths in order to retrieve the rotation rate and explore the
presence of active weather.

We have shown in this paper that, if the rotation period of an
Earth-like planet can be determined accurately, one can then fold
the photometric light curves at the rotation period to study
regional properties of the planet's surface and/or atmosphere. Most
significantly we could learn if dynamic weather is present on an
Earth-like exoplanet, from deviations from a fixed phase curve.  In
contrast, a cloud-free planet with continents and oceans would not
show such light curve deviations. With phased light curves we could
study local surface or atmospheric properties with follow-up
photometry, spectroscopy, and polarimetry, to detect surface and
atmospheric inhomogeneities and to improve the sensitivity to
localized biomarkers. Finally, we have also provided guidance for
the necessary specifications for future space missions.

\acknowledgments

Research by E. Pall\'e was supported by a `Ramon y Cajal'
fellowship. Support for E.B. Ford was provided by NASA through
Hubble Fellowship grant HST-HF-01195.01A awarded by the Space
Telescope Science Institute, which is operated by the Association of
Universities for Research in Astronomy, Inc., for NASA, under
contract NAS 5-26555.

\clearpage

\begin{table}
\caption{The existence of 21 years of global cloud observations
allow us to simulate the photometric time series of Earth for each
of these years, with the exact same geometrical configurations, with
only clouds changing. For each year we have calculated the main
periodicities resulting from the photometric simulations analysis.
In the table, we show the percentage of years in which the main
periodicity is found to be (i) $24\pm \Delta$ hours, (ii) $12\pm
\Delta$ hours or (iii) other periods. For example a $95\%$ value in
a given periodicity, means that for 20 of the 21 available years
(cloud configurations) that was the primary periodicity. Fourier and
autocorrelation analysis results are both shown. In all cases $\pm
\Delta$ is taken as the exposure time.} \vspace{2mm}
\begin{tabular}{|l|ccc|ccc|}
\hline
Area  & 24 & 12 & other  & 24 & 12 & other \\
\hline
  & \multicolumn{3}{c}{Fourier} & \multicolumn{3}{c}{Auto} \\
\hline
  & \multicolumn{6}{c}{S/N=20, Exp=0.5h, Followup=2w } \\
\hline
N. Pole &           38 &            4 &           58 &          100 &        0 &            0 \\
Lat +45 &            9 &           76 &           15 &          100 &        0 &            0 \\
Equator &        0 &           71 &           29 &          100 &        0 &            0 \\
Lat -45 &            4 &           57 &           39 &          100 &        0 &            0 \\
S. Pole &            9 &           23 &           68 &          100 &        0 &            0 \\
\hline
  & \multicolumn{6}{c}{S/N=20, Exp=0.5h, Followup=8w } \\
\hline
N. Pole &           66 &            4 &           30 &          100 &        0 &            0 \\
Lat +45 &            4 &           71 &           25 &          100 &        0 &            0 \\
Equator &        0 &           80 &           20 &          100 &        0 &            0 \\
Lat -45 &        0 &           76 &           24 &          100 &        0 &            0 \\
S. Pole &           23 &           42 &           35 &          100 &        0 &            0 \\
\hline
  & \multicolumn{6}{c}{S/N=5, Exp=0.5h, Followup=2w } \\
\hline
N. Pole &            4 &        0 &           96 &           33 &        0 &           67 \\
Lat +45 &        0 &            9 &           91 &           52 &        0 &           48 \\
Equator &        0 &           33 &           67 &           76 &        0 &           24 \\
Lat -45 &        0 &           28 &           72 &           80 &        0 &           20 \\
S. Pole &        0 &        0 &          100 &           23 &        0 &           77 \\
\hline
  & \multicolumn{6}{c}{S/N=20, Exp=1.7h, Followup=2w } \\
\hline
N. Pole &           85 &            4 &           11 &           80 &        0 &           20 \\
Lat +45 &           66 &           28 &            6 &           85 &        0 &           15 \\
Equator &           38 &           47 &           15 &           85 &        0 &           15 \\
Lat -45 &           57 &           33 &           10 &           90 &        0 &           10 \\
S. Pole &           71 &            9 &           20 &           80 &        0 &           20 \\
\hline
\end{tabular}
\label{tab1}
\end{table}

\begin{table}
\caption{Same as Table~1, but the calculations are done for a
cloud-free Earth.} \vspace{2mm}
\begin{tabular}{|l|ccc|ccc|}
\hline
Area  & 24 & 12 & other  & 24 & 12 & other \\
\hline
  & \multicolumn{3}{c}{Fourier} & \multicolumn{3}{c}{Auto} \\
\hline
  & \multicolumn{6}{c}{S/N=20, Exp=0.5h, Followup=2w } \\
\hline
N. Pole &           95 &            4 &            1 &          100 &        0 &            0 \\
Lat +45 &           19 &           80 &            1 &          100 &        0 &            0 \\
Equator &            4 &           95 &            1 &          100 &        0 &            0 \\
Lat -45 &            4 &           95 &            1 &          100 &        0 &            0 \\
S. Pole &        0 &          100 &            0 &          100 &        0 &            0 \\
\hline
  & \multicolumn{6}{c}{S/N=20, Exp=0.5h, Followup=8w } \\
\hline
N. Pole &           95 &            4 &            1 &        0 &        0 &          100 \\
Lat +45 &            4 &           95 &            1 &            4 &        0 &           96 \\
Equator &        0 &          100 &            0 &          100 &        0 &            0 \\
Lat -45 &            4 &           95 &            1 &          100 &        0 &            0 \\
S. Pole &        0 &          100 &            0 &           95 &        0 &            5 \\
\hline
  & \multicolumn{6}{c}{S/N=5, Exp=0.5h, Followup=2w } \\
\hline
N. Pole &           52 &        0 &           48 &          100 &        0 &            0 \\
Lat +45 &           28 &           71 &            1 &          100 &        0 &            0 \\
Equator &            4 &           95 &            1 &          100 &        0 &            0 \\
Lat -45 &            4 &           95 &            1 &          100 &        0 &            0 \\
S. Pole &        0 &           95 &            5 &          100 &        0 &            0 \\
  & \multicolumn{6}{c}{S/N=20, Exp=1.7h, Followup=2w } \\
\hline
N. Pole &          100 &        0 &            0 &          100 &        0 &            0 \\
Lat +45 &          100 &        0 &            0 &          100 &        0 &            0 \\
Equator &            4 &           95 &            1 &          100 &        0 &            0 \\
Lat -45 &            9 &           90 &            1 &          100 &        0 &            0 \\
S. Pole &           76 &           23 &            1 &          100 &        0 &            0 \\
\hline
\end{tabular}
\label{tab2}
\end{table}

\clearpage

\end{document}